\documentclass[epsfig,12pt]{article}
\usepackage{epsfig}
\usepackage{graphicx}

\newcommand{\beq}{\begin{equation}}   
\newcommand{\eeq}{\end{equation}}
\newcommand{\beqn}{\begin{eqnarray}}   
\newcommand{\eeqn}{\end{eqnarray}}

\newcommand{\gsim}{\lower.7ex\hbox{$
\;\stackrel{\textstyle>}{\sim}\;$}}
\newcommand{\lsim}{\lower.7ex\hbox{$
\;\stackrel{\textstyle<}{\sim}\;$}}

\begin{document}

\begin{titlepage}

\begin{flushright}
FTPI-MINN-12/40, UMN-TH-3131/12\\
\end{flushright}

\vspace{0.7cm}

\begin{center}
{  \large \bf  Simple Models with Non-Abelian Moduli on \\[3mm]Topological Defects}
\end{center}
\vspace{0.6cm}

\begin{center}
 {\large 
    M. Shifman}

\end {center}

\begin{center}

{\em William I. Fine Theoretical Physics Institute, University of Minnesota,
Minneapolis, MN 55455, USA} 
\end {center}

\vspace{2cm}

\begin{center}
{\large\bf Abstract}
\end{center}

\hspace{0.3cm}
	I explain  how conventional topological defects -- the Abrikosov-Nielsen-Olesen (ANO) string and domain walls --
	can acquire non-Abelian moduli localized on their world sheets. The set-up is conceptually similar and generalizes that used by Witten for cosmic strings \cite{W}.
	
\vspace{2cm}

\end{titlepage}

\newpage

\section{Introduction}

The discovery of non-Abelian strings \cite{AHDT} (i.e. those with non-Abelian moduli fields on the string world sheet)
paved the way to many applications, both in supersymmetric \cite{book} and non-supersymmetric theories  \cite{Gorsky}. The models supporting strings with non-Abelian moduli which are discussed in the literature are rather advanced, especially in the supersymmetric case. At the same time the physical essence of the phenomenon is quite simple. Here I explain the occurrence of the non-Abelian moduli
in a simple set-up. All we need is: (i)
 the bulk theory having an unbroken {\em global} non-Abelian symmetry  
$G$ and supporting topological defects (e.g. strings
or domain walls); and (ii) the breaking of above global symmetry $G$  down to a {\em global} subgroup $H$ on the given defect.
These two requirements can be easily implemented.
Then, at the classical level we will have $\nu= $ dim$\,G-$dim$\,H$ zero modes of an ``orientational" type,
localized on the given topological defect, in addition to familiar translational modes (which as a rule are decoupled from orientational \cite{LSV}). At low energies the corresponding moduli fields are described by a sigma model with the $G/H$ coset space as the target space. Quantization of the low-energy $G/H$ sigma model on the world sheet of the
topological defect under consideration may or may not lift the zero modes.

I will assume weak coupling justifying quasiclassical treatment. A few sample models to be considered below (and similar)
may play a role of the Ginzburg-Landau models in future applications. The theoretical set up to be presented below  is conceptually similar and generalizes that used by Witten for cosmic strings \cite{W}. Note that some generalizations of Witten's construction were considered in the 1980s and early 90s \cite{WG}. They do not overlap with the non-Abelian moduli construction I will discuss. Of more relevance is the recent publication \cite{lil} in which non-Abelian bosonic currents in cosmic strings were discussed.

\section{Non-Abelian string from Abrikosov-\\
Nielsen-Olesen vortex}
\label{nas}

First, I will briefly outline the construction of the ANO string which has no moduli on its world sheet other than
translational (see e.g. \cite{shi}). The ANO string
is a soliton in the  U(1) gauge theory 
with a single charged scalar field whose vacuum expectation value breaks U(1)
spontaneously. 
The model is described by the 
Lagrangian
 \beq
{\cal L}_{\rm v} = -\frac{1}{4e^2}F_{\mu\nu}^2 + \left| {\mathcal D}^\mu\phi\right|^2
  -V(\phi )\, 
  \label{tpi16}
\eeq
where 
\beq
F_{\mu\nu} = \partial_\mu A_\nu -\partial_\nu A_\mu\,,\qquad {\mathcal D}_\mu\phi = (\partial_\mu -iA_\mu )\phi\,.
\eeq
The potential energy $V(\phi )$ must be chosen in such a way as to ensure Higgsing
of U(1) in the bulk, 
\beq
V= \lambda \left(|\phi |^2 -v^2
\right)^2\,,
\eeq
where $v$ is assumed to be real and positive (no loss of generality).
In the vacuum in the unitary gauge  
\beq
A_\mu = 0,\qquad \phi = v\,.
\label{622one}
\eeq
The U(1) photon is Higgsed and  
acquires the mass
\beq
m_\gamma= \sqrt 2 e v\,,
\label{l710}
\eeq
${\rm Im}\,\phi$ is eaten by the Higgs mechanism, while
Re$\,\phi(x)=v+\eta(x)/\sqrt{2}$, where
the  real scalar field $\eta(x)$   is not eaten up by the photon.
 Its mass is
\beq
m_\eta= 2 \sqrt { \lambda}\,  v\,.
\label{l711}
\eeq
We will assume that $m_\eta > m_\gamma$, but not much larger, i.e. $m_\eta \sim m_\gamma$.
This is not crucial, though.

Now, as well-known, this model supports topologically stable vortices (strings).
Indeed,
let us first consider all non-singular field configurations that are static
(time-independent) in the gauge $A_0 =0$. Then the energy functional takes the form
\begin{eqnarray}
{\cal E} [\vec A(\vec x), \phi (\vec x )] &=& \int dz\,\int d^2 x
\left[\frac{1}{4e^2} F_{ij}F_{ij} +\left| {\mathcal D}_i\phi \right|^2 +V(\phi )
\right]
\nonumber \\[2mm]
&=& 
L\times \int d^2 x
\left[\frac{1}{4e^2} F_{ij}F_{ij} +\left| {\mathcal D}_i\phi \right|^2 +V(\phi )
\right],
 \label{wedtwop}
\end{eqnarray}
where $L\to \infty$ is the string length (it is assumed to be oriented along the $z$ axis),
while the integral in the second line presents the string tension $T$.
Requiring $T$ to be finite we observe that
  $V(\phi )\to 0$ at $|\vec x_\perp |\to\infty$, i.e.
\beq
|\phi | \to v \quad\mbox{at}\quad |\vec x_\perp |\to\infty\,.
\eeq
Let us choose a circle of large radius $R$ (eventually $R \to\infty $)
centered at $x=y=0$. The absolute value of $\phi$ on this circle must be $v$,
however, the phase of the field $\phi$ is not fixed by the condition  
 $\int d^2x\, V(\vec x_\perp)<\infty$. The minimal ANO string is obtained if  
\beq
\phi = v\, e^{i\alpha }
\label{otlc}
\eeq
on the large circle. Here $\alpha$ is the polar angle in the $x,y$ plane. 

The {\em ansatz} (\ref{otlc}) combined with the requirement of finiteness 
of the kinetic term of the $\phi$ field generates a gauge potential in the perpendicular plane,
$A_\perp \to A_i$ $(i=1,2)$. At large distances from the string axis $A_i$ is pure gauge,
\beq
A_i=   \partial_i\alpha = -\varepsilon_{ij}
\frac{x_j}{r^2}\,,\quad i,j=1,2\,,
\label{wedthirteen}
\eeq
where $\varepsilon_{ij}$ is the two-dimensional Levi-Civita tensor.\index{Levi--Civita tensor}
It is clear that then both ${\mathcal D}_i\phi$ and $F_{ij}$ fall off at infinity faster
than
$1/r^2$ (in fact, they fall off exponentially fast), and the  integral for $T$ converges,
\beq
T\sim v^2\,.
\eeq
It is easy to see that this vortex carries a (quantized) magnetic flux in the $z$ direction.
It has two zero modes (translational modes in the $x$ and $y$ directions) which become sterile moduli fields on the
 string world sheet.

The full elementary vortex solution is
parametrized by
two profile functions $\varphi(r)$ and  $f(r)$ as follows: 
\begin{equation}
\phi(x) = v\,\varphi(r) e^{i\,\alpha }\;,\qquad
A_i(x) =-\frac{1}{n_e }\varepsilon_{ij}\,\frac{x_j}{r^2}\, \left[1- f(r)\right]
\, ,
\label{profil}
\end{equation}
where $r=|x_\perp |=\sqrt{x^2+y^2}$ is the distance from the string axis
and $\alpha $ is the polar
angle, as above.
 
The boundary conditions for the   profile functions
are rather obvious from the form of the  {\em ansatz} (\ref{profil}).
 At large distances
\beq
\varphi (\infty)= 1\,,\quad f(\infty)= 0\,.
\label{624four}
\eeq
The absence of singularities at the origin  requires
\begin{equation}
\varphi(0)=0\,,\quad f(0)=1\,.
\label{624five}
\end{equation}
Thus, in the vortex core the $\phi$ expectation value vanishes and the original U(1) 
gauge symmetry is restored. The transverse size of the ANO flux tube $\sim 1/m_{\eta,\gamma}$.

So far, this an absolutely standard construction. Now I will extend it. Introduce a triplet field $\chi^i$ (here $i=1,2,3$)
endowed with a (globally) SO(3) invariant interaction,
\beqn
{\cal L}_\chi &=& \partial_\mu \chi^i \, \partial^\mu \chi^i - U(\chi, \phi)\,,
\label{14}\\[2mm]
U &=&  \gamma\left[\left(-\mu^2 +|\phi |^2
\right)\chi^i \chi^i + \beta \left( \chi^i \chi^i\right)^2\right],
\label{15}
\eeqn
where $\gamma$ is a (positive) coupling constant. 
For simplicity I will assume  $\beta >1$
and the field $\chi$ to be real. The parameter $\mu$ is real and positive, with the condition
\beq
\mu < v\,,
\label{12mv}
\eeq
but not much smaller. For the validity of our consideration we must require

\beq
\gamma \mu^2 \gg \lambda v^2\,,
\eeq
so that the length scale of variation of the $\eta ,\,\,\gamma$ fields is larger than that of the $\chi$ fields.

In the bulk the expectation value of $\phi$ does not vanish, $|\phi |=v$. Equation (\ref{15}) implies then that
$\chi $ is stable, no vacuum condensate of $\chi$ develops, and the global O(3) symmetry remains unbroken.
At the same time,
in the vortex core $\phi $ vanishes  destabilizing the $\chi$ field which develops an expectation value, 
\beq
\chi^2 = \frac{\mu^2}{2\beta}\,,
\label{19}
\eeq
implying, in turn, that in the core the O(3) symmetry is spontaneously broken. 
While the absolute value of $\chi^i$ is fixed at $|\chi^i_{*}| =\mu/\sqrt{ 2\beta}$ by the energy minimization
the angular orientation of the  vector $\chi^i$ is arbitrary. Independently of
$\chi^i_{*}$, the pattern of the symmetry breaking in the string core is
\beq
O(3) \to O(2)\,.
\label{20}
\eeq
Correspondingly, the vortex solution (more exactly, its $\chi$ component)
will depend on two moduli whose dynamics is determined
by the $O(3)/O(2)=CP(1)$ coset. Differentiating the solution with respect to these two collective coordinates
we get the explicit form of the zero modes. The low-energy theory on the string world sheet
is the $CP(1)$ model for the orientational moduli fields (in addition to two decoupled
translational moduli fields),
namely,
\beq
{\mathcal L}_{\rm sws} = \frac{1}{\beta} \, \left(\partial^a n^i\right)\left(\partial_a n^i\right),\qquad n^in^i=1\,,\qquad
a=0,3\,.
\label{21}
\eeq
The subscript sws means string world sheet.

Classically and in perturbation theory the above moduli fields are massless. However, from the
Coleman theorem we know that massless non-sterile boson fields cannot exist in two dimensions \cite{coleman}.
And indeed, the exact solution of the $CP(1)$ model (which is asymptotically free in the UV, but 
strongly coupled in the IR \cite{polyakov}) exhibits a mass gap generation and complete restoration of O(3). 
The would-be Nambu-Goldstone (NG) bosons on the string world sheet become quasi-NG bosons,
provided the mass scale $\Lambda$ which is nonperturbatively generated in CP(1) 
is small,
\beq
\Lambda \ll v\,.
\eeq
If $\beta\gg 1$ the above condition is met.

\section{Non-Abelian moduli fields on domain walls}
\label{nam}

The general idea is the same as in Sect. \ref{nas}: an unbroken global symmetry in the vacuum
combined  with a domain wall which breaks a part of the above global symmetry in its core.
As a pedagogical example we will consider the same set-up (\ref{14}) and (\ref{15}) in conjunction with the simplest model of the complex field $\phi$ supporting an appropriate domain wall.

Such a model is given by the Lagrangian   (see e.g. \cite{shi})
\begin{equation}
{\mathcal L}_{\rm w} = (\partial^{\mu} \phi^\dagger)(\partial_{\mu} \phi) - V(\phi, \phi^\dagger)\,, 
\label{wzcompx2}
\end{equation}
where
\begin{equation}
V(\phi , \bar\phi) = \left| \frac{\partial \,{ W}(\phi)}{\partial 
\phi}\right|^2\,,\qquad { W}(\phi) = \frac{m^2}{\lambda} \,\phi - \frac{\lambda}{3}\, \phi^3\,,
\label{spotc}
\end{equation}
and the constants $m$ and $\lambda$ are assumed to be real and positive.

The potential~(\ref{spotc}) implies two degenerate classical
vacua,
\begin{equation}
\phi_{*} = \pm v\, , \qquad v\equiv \frac{m}{\lambda} \,.
\label{vacua}
\end{equation}
 Both vacua are physically equivalent. This equivalence
 could be explained by the spontaneous 
breaking of the
$Z_2$ symmetry, $\phi \to -\phi$, present in the 
action.
The static field configuration interpolating between the two degenerate vacua is  the 
domain wall (which   is 
topologically stable).
 Assume for  definiteness that the wall lies in the $xy$ plane. 
The wall tension $T_{\rm w}$  
(the energy per unit area $T_{\rm w} = E_{\rm w}/{A}$) is 
\beq
T= \frac{8m^3}{3\lambda^2}\,.
\eeq
while the wall solution takes the form
\begin{equation}
\phi_{\rm w} = \frac{m}{\lambda}\tanh (mz)\, ,
\label{wallsol}
\end{equation}
implying that the wall thickness is $\sim m^{-1}$.

Now we combine ${\mathcal L}_{\rm w}$ from Eq. (\ref{wzcompx2}) with ${\mathcal L}_{\chi}$ from Eqs. 
(\ref{14}) and (\ref{15}). In much the same way as in Section \ref{nas},
in the bulk the $\phi$ condensate,  $|\phi_{*}|^2=v^2$, guarantees stability of the
$\chi$ field. The global O(3) symmetry is unbroken.

At the same time, in the wall core where $\phi$ is close to zero (see Eq. (\ref{wallsol}) at $z$ close to zero)
the $\chi$ field becomes unstable and develops an expectation value, see Eq. (\ref{19}). Thus, in the wall core
the global symmetry breaking (\ref{20}) occurs, giving rise to massless $O(3)/O(2)$ moduli 
described by  the Lagrangian (\ref{21}) on the wall world sheet.

There are two distinction, however. First the index $a$ now runs over $a=0,1,2$, i.e. the world sheet theory is three-dimensional. This distinction implies, in turn, that the world sheet theory is admittedly non-renormalizable low-energy effective theory and, as a result, less infrared-dependent. There is no reason to expect that the classical
masslessness of the $n^i$ fields  will be lifted at the quantum level. In other words,
the O(3) symmetry of the bulk will not be restored on the wall implying that the
moduli fields will remain  
exactly massless
NG bosons localized on the wall.  Needless to say, in addition to the $n^i$ fields, one (sterile) translational NG field
is localized on the wall too. The corresponding sigma model will have a factorized structure with no coupling between the
transnational and orientational moduli fields. Such factorization is common \cite{LSV}.

\section{Other patterns}
\label{op}

The same strategy as above can be easily used to obtain a variety of non-Abelian moduli localized on topological
defects, such as domain walls or strings. Here I will give an extra example.

Instead of the O(3) triplet $\chi^i$ field as in (\ref{14}) and (\ref{15}) let us consider an SU(2) doublet $\eta^p$
of  complex scalar fields
($p=1,2$). We will couple this doublet to ${\mathcal L}_{\rm w}$ analogously to (\ref{14}) and (\ref{15}),
\beqn
{\cal L}_\eta &=& \partial_\mu \bar \eta_i \, \partial^\mu \eta^i - U_\eta(\bar\eta, \eta, \phi)\,,
\label{28}\\[2mm]
U_\eta &=&  \gamma\left[\left(-\mu^2 +|\phi |^2
\right)\bar\eta_p \eta^p + \beta \left( \bar\eta_p \eta^p\right)^2\right],
\label{29}
\eeqn
It is obvious that this theory has a global $SU(2)$ symmetry, by construction. Actually, it has an
$SU(2)\times SU(2) $ symmetry \cite{Shiv} due to the fact that $SU(2)$ is a quasi-real group. To see that this is indeed the case
let us introduce a two-by-two matrix $X$,
\beq
X = \left(
\begin{array}{cc}
\eta^1  &  -(\eta^2)^* \\[1mm]
\eta^2  & (\eta^1)^*
\end{array}
\right).
\eeq
In terms of $X$ the Lagrangian (\ref{28}), (\ref{29}) takes the form \cite{Shiv}
\beq
{\mathcal L} = \frac{1}{2}{\rm Tr}
\left(  \partial^\mu X  \right)^\dagger \left(\partial_\mu X\right)
-\gamma
\left[ \left(-\mu^2 +|\phi |^2
\right)\frac{1}{2}\, {\rm Tr} \, X^\dagger X + \frac{\beta}{4}\left( {\rm Tr} \, X^\dagger X\right)^2
\right].
\eeq
This Lagrangian  is obviously invariant under the transformation
\beq
X(x) \to U X(x) M^{-1}\,,
\label{32}
\eeq
 where $U$ and $M$ are arbitrary $x$-independent matrices from 
${\rm SU}(2)$.
The $SU(2)\times SU(2) $ symmetry  is apparent. 
In the vacuum where $|\phi_*| > \mu$ the expectation value of $X$ vanishes.
 The full $SU(2)\times SU(2) $ symmetry is unbroken.
If, however, $|\phi_*| \sim 0$
the value of $X$ approaches
\beq
X_{*} = \frac{\mu}{\sqrt {2\beta}} \left(
\begin{array}{cc}
1  &  0 \\[1mm]
0  & 1
\end{array}
\right)
\label{33}
\eeq
(corresponding to $\eta^1 =1,\,\,\eta^2=0$).
This vacuum expectation value breaks ${ SU}(2)_{L}$ and $ { SU}(2)_{ R}$,
but the diagonal global SU(2) corresponding to $U=M$ in (\ref{32}) remains unbroken. 

Now, we combine this Lagrangian with (\ref{wzcompx2}). Repeating the argumentation of the previous sections
we expect that in the core of the wall the global symmetry is broken down to SU(2), with three NG modes
in the coset
\beq
\frac{SU(2)\times SU(2)}{SU(2)}\,.
\eeq
In other words, if a solution with (\ref{33}) in the core exists, there should exist a family of solutions with the unit matrix replaced by
$U$, an arbitrary two-by-two matrix from SU(2). These matrices are parametrized by three parameters (moduli). When they are endowed by the world sheet coordinate dependence, they become moduli fields localized on the
topological defect (i.e. the moduli fields  in 1+2 dimensions, if localized on the wall, and in 1+1 dimension, if localized on the vortex).

This is the standard chiral model for pions, whose Lagrangian can be written as
\beq
{\mathcal L} = \frac{1}{4g^2}\,{\rm Tr}\, \partial^\mu U\partial_\mu U^\dagger\,. 
\eeq
In 1+1 dimensions this model is renormalizable and, moreover, asymptotically free
  \cite{Leutwyler:1991tv}. A mass gap is expected  to be generated at strong coupling, due to infrared interactions. In 1+2  dimensions this model is non-renormalizable; no mass gap generation is expected.

\section{Conclusions}

A few simple models discussed 
above demonstrate the fact that the occurrence of non-Abelian moduli
localized on topological defects is a common and simple phenomenon, rather than 
an exotic and rare possibility.

\vspace{5mm}
I am grateful to A. Yung for comments.
This work is supported in part by DOE grant DE-FG02- 94ER-40823.

\end{document}